\documentclass[10pt,letterpaper]{article}
\usepackage[top=0.85in,left=2.75in,footskip=0.75in]{geometry}

\usepackage{amsmath,amssymb}

\usepackage{changepage}

\usepackage{textcomp,marvosym}

\usepackage{cite}

\usepackage{nameref,hyperref}

\usepackage[right]{lineno}

\usepackage[nopatch=eqnum]{microtype}
\DisableLigatures[f]{encoding = *, family = * }

\usepackage[table]{xcolor}

\usepackage{array}
\usepackage{textcomp,marvosym}
\usepackage[utf8]{inputenc}
\usepackage[T1]{fontenc}
\usepackage{float}
\usepackage{algorithm}
\usepackage{algpseudocode}
\usepackage{float}
\usepackage{tabularx}
\usepackage{booktabs}
\usepackage{placeins}
\usepackage{cite}
\newcolumntype{+}{!{\vrule width 2pt}}

\newlength\savedwidth

\raggedright
\usepackage[aboveskip=1pt,labelfont=bf,labelsep=period,justification=raggedright,singlelinecheck=off]{caption}

\makeatletter
\renewcommand{\@biblabel}[1]{\quad#1.}
\makeatother

\usepackage{lastpage,fancyhdr,graphicx}
\usepackage{epstopdf}
\fancyheadoffset[L]{2.25in}
\fancyfootoffset[L]{2.25in}
\begin{document}
\vspace*{0.2in}

\begin{flushleft}
{\Large
\textbf\newline{Machine Learning-Based Prediction of Mortality in Geriatric Traumatic Brain Injury Patients} 
}
\newline
\\

Yong Si\textsuperscript{1},
Junyi Fan\textsuperscript{1},
Li Sun\textsuperscript{1},
Shuheng Chen\textsuperscript{1},
Elham Pishgar\textsuperscript{2},
Kamiar Alaei\textsuperscript{3},
Greg Placencia\textsuperscript{4},
Maryam Pishgar\textsuperscript{1*}
\textsuperscript{\textpilcrow}
\\
\bigskip
\textbf{1} Department of Industrial and Systems Engineering, University of Southern California, 3715 McClintock Ave GER 240, Los Angeles, 90087, California, United States
\\
\textbf{2} Colorectal Research Center, Iran University of Medical Sciences, Tehran Hemat Highway next to Milad Tower, Tehran, 14535, Iran
\\
\textbf{3} Department of Health Science, California State University, Long Beach (CSULB), 1250 Bellflower Blvd, Long Beach, 90840, California, United States
\\
\textbf{4} Department of Industrial and Manufacturing Engineering, California State Polytechnic University, Pomona, 3801 W Temple Ave, Pomona, 91768, California, United States
\\
\bigskip

%
%




\textpilcrow Membership list can be found in the Acknowledgments section.

* pishgar@usc.edu

\end{flushleft}
\section*{Abstract}
Traumatic Brain Injury (TBI) is a major contributor to mortality among older adults, with geriatric patients facing disproportionately high risk due to age-related physiological vulnerability and comorbidities. Early and accurate prediction of mortality is essential for guiding clinical decision-making and optimizing ICU resource allocation. In this study, we utilized the MIMIC-III database to identify geriatric TBI patients and applied a machine learning framework to develop a 30-day mortality prediction model. A rigorous preprocessing pipeline—including Random Forest-based imputation, feature engineering, and hybrid selection—was implemented to refine predictors from 69 to 9 clinically meaningful variables. CatBoost emerged as the top-performing model, achieving an AUROC of 0.867 (95\% CI: 0.809–0.922), surpassing traditional scoring systems. SHAP analysis confirmed the importance of GCS score, oxygen saturation, and prothrombin time as dominant predictors. These findings highlight the value of interpretable machine learning tools for early mortality risk stratification in elderly TBI patients and provide a foundation for future clinical integration to support high-stakes decision-making in critical care.



\section*{Introduction}
Traumatic Brain Injury (TBI) arises from external mechanical forces—such as blunt trauma, acceleration-deceleration, or penetrating injuries—that disrupt neurological function\cite{corrigan2010epidemiology,mckee2015neuropathology,ng2019traumatic}. Its pathophysiology involves primary structural damage and secondary cascades (e.g., neuroinflammation, axonal injury), driving heterogeneous clinical outcomes\cite{hume2013biomechanics,ng2019traumatic,de2024pathophysiology}. Globally, TBI affects millions annually, with an estimated 69 million cases each year, making it a leading cause of disability and death\cite{dewan2018estimating,injury2019global,hyder2007impact}. The economic burden of TBI is substantial, with annual healthcare costs in the U.S. alone exceeding \$76 billion, encompassing acute care, rehabilitation, and lost productivity\cite{taylor2017traumatic,finkelstein2006incidence}. 

Older adults are disproportionately vulnerable to TBI, with falls—responsible for over 50\% of geriatric TBI cases—driven by age-related risks such as polypharmacy, gait instability, and osteoporosis\cite{thomas2008fall,cusimano2020population,harvey2012traumatic}.
In the U.S., annual TBI-related healthcare costs for older adults exceed \$10 billion, reflecting acute care, rehabilitation, and long-term disability \cite{thompson2012utilization,florence2018medical}. Especially, older adults with TBI face disproportionately high mortality rates, with studies reporting 2–3-fold increased risk of death compared to younger cohorts, driven by complications such as sepsis, respiratory failure, and neurodegenerative exacerbations\cite{mcintyre2013mortality}.
Given the global aging population—projected to double by 2050—this mortality burden is poised to escalate, straining healthcare systems \cite{grinin2023global}.  Considering this urgent trend, accurate mortality prediction tools (e.g., machine learning-driven models or biomarker panels) are now critical to identify high-risk patients, tailor interventions (e.g., early palliative care), and prioritize resource allocation. Concurrently, age-specific prevention (e.g., fall reduction programs) and rapid diagnostic protocols remain essential to mitigate TBI incidence and improve outcomes in this vulnerable population\cite{maas2015tbi}.

Fu et al. (2017)\cite{fu2017predictors} analyzed trends in TBI hospitalizations and mortality among elderly adults (65+) in Canada from 2006 to 2011 using a population-based database. Advanced age, comorbidities, and injury severity were independent predictors of both falls and in-hospital mortality. The authors suggest that prevention efforts should focus on the "older old" (85+) and those with multiple comorbidities, and recommend that healthcare facilities be prepared to manage this growing, complex patient population.

Bobeff et al. (2019) developed the Elderly Traumatic Brain Injury Score (eTBI Score) to predict 30-day mortality or vegetative state in geriatric TBI patients. The study analyzed data from 214 patients aged $\geq$ 65 years, focusing on demographics, medical history, and clinical factors. Key predictors identified through logistic regression included Glasgow Coma Scale (GCS) motor score (OR 0.17), comorbid cardiac, pulmonary, or renal dysfunction or malignancy (OR 2.86), platelet count $\leq$ 100 × 10$^9$ cells/L (OR 13.60), and red blood cell distribution width $\geq$ 14.5\% (OR 2.91). The eTBI Score provides a practical tool for clinical decision-making and risk stratification in elderly TBI patients, offering reliable outcomes for managing treatment\cite{bobeff2019mortality}.

Huang et al. (2024) assessed the utility of the Geriatric Trauma Outcome Score (GTOS) in predicting mortality in older adults with isolated moderate to severe TBI. The study included 5,543 patients and found that higher GTOS was significantly associated with increased mortality, with the optimal cutoff value for mortality prediction identified as 121.5 (AUC = 0.813). Patients with GTOS $\geq$ 121.5 had higher odds of death (OR 2.64; 95\% CI 1.93–3.61) and longer hospital stays. These findings suggest that GTOS is an effective tool for risk stratification in TBI patients, though further refinement is needed for broader clinical application\cite{huang2024geriatric}.

While traditional models such as eTBI and GTOS have been commonly used to predict outcomes in geriatric TBI patients, they have certain limitations. These models typically focus on a narrow range of clinical factors and may not fully account for the complexities of older patients, such as comorbidities and frailty. Furthermore, they often struggle with incorporating high-dimensional data, such as laboratory results and imaging findings, which can reduce their accuracy in predicting mortality. Consequently, there is a critical need for a more comprehensive and adaptable predictive tool that can improve the accuracy and reliability of mortality predictions in elderly TBI patients, addressing the shortcomings of existing models.

In recent years, machine learning techniques have gained significant traction in predicting clinical outcomes, with the CatBoost algorithm emerging as a highly effective tool. CatBoost offers several key advantages for medical data analysis: (1) it incorporates robust regularization mechanisms that help prevent overfitting, particularly in datasets with high-dimensional features, (2) it efficiently handles missing values, a common challenge in clinical data, and (3) it provides superior feature importance quantification, which enhances the interpretability of the model for clinical decision-making. These characteristics make CatBoost a promising approach for clinical outcome prediction\cite{hancock2020catboost}. For example, Li et al. (2023) developed a CatBoost-based model to predict hospital mortality in ICU patients receiving mechanical ventilation (MV) using the MIMIC-III database. The model showed strong discriminative performance, achieving an AUROC of 0.862 (95\% Confidence Interval (CI): 0.850–0.874) in internal validation, surpassing the best reported AUROC of 0.821. Additionally, the model demonstrated improved accuracy (0.789), F1-score (0.747), and calibration, outperforming other machine learning models, including XGBoost, Random Forest, and Support Vector Machine (SVM)\cite{li2024machine}. Safaei et al. (2022) developed an optimized CatBoost-based model, E-CatBoost, to predict ICU mortality status upon discharge using data from the first 24 hours of admission. The model was trained and validated using the eICU-CRD v2.0 dataset, which includes over 200,000 ICU admissions. The E-CatBoost model achieved AUROC scores ranging from 0.86 to 0.92  across twelve disease groups, outperforming baseline models by 7 to 18 percent. The study identified key features in mortality prediction, including age, heart rate, respiratory rate, blood urea nitrogen, and creatinine levels, providing valuable insights into critical patient conditions for early intervention\cite{safaei2022catboost}. These successful cases in developing machine learning-based models for clinical outcome prediction illustrate a feasible and promising future for this approach.

This study introduces several key innovations that enhance the predictive performance and clinical applicability of machine learning models for mortality prediction in geriatric TBI patients.

\begin{itemize}
    \item Hybrid Feature Selection Strategy: The study utilized a combination of Random Forest-based importance and Recursive Feature Elimination (RFE) to refine an initial pool of candidate variables, resulting in a set of clinically relevant features. This method effectively reduced dimensionality and ensured the retention of features most critical for mortality prediction.
\item Data Imputation and Preprocessing: To address common challenges in clinical datasets, such as missing data and heterogeneity in measurements, the study implemented Random Forest-based imputation. This approach preserved the interdependencies of variables, enhancing the robustness of the model.

\item Model Selection and Evaluation: A range of machine learning models, including CatBoost, LightGBM, and XGBoost, were evaluated. CatBoost demonstrated the best performance, achieving an AUROC of 0.867 (95\% CI: 0.809–0.922), outperforming other models such as LightGBM and XGBoost, which had AUROCs of 0.852 and 0.855, respectively. This indicates that CatBoost provided the most reliable predictions for 30-day mortality in geriatric TBI patients, with good balance between sensitivity and specificity.

\item Interpretability and Clinical Integration: To ensure clinical transparency and utility, SHAP (SHapley Additive exPlanations) analysis was applied to interpret the feature contributions. Key predictors such as GCS score, respiratory rate, and prothrombin time were identified as major contributors to the model’s predictions, providing actionable insights for clinicians.
\end{itemize}



\section*{Methods}
\subsection*{Data Source and study design}

This study utilized the Medical Information Mart for Intensive Care III (MIMIC-III) database, a publicly available and extensively curated critical care dataset developed by the Laboratory for Computational Physiology at the Massachusetts Institute of Technology (MIT). MIMIC-III comprises de-identified health records of over 40,000 patients admitted to the Beth Israel Deaconess Medical Center (BIDMC) between 2001 and 2012. The database complies with the Health Insurance Portability and Accountability Act (HIPAA) through systematic de-identification procedures, including date shifting and removal of personally identifiable information.

MIMIC-III offers a wide range of structured and unstructured clinical data, including patient demographics, clinical notes, physiological time series, laboratory results, medication administration, and treatment information \cite{johnson2016mimiciii}.

This study implemented a structured and clinically grounded machine learning pipeline to predict 30-day mortality in geriatric TBI patients. The workflow integrated rigorous data selection, preprocessing, feature engineering, feature selection, model development, and statistical validation, as illustrated in Algorithm~\ref{alg:tbi_mortality}.

\begin{algorithm}[H]
\caption{\textbf{ML Pipeline for 30-Day Mortality Prediction in Geriatric TBI Patients}}
\label{alg:tbi_mortality}
\begin{algorithmic}[1]
\Require MIMIC-III ICU data with confirmed TBI diagnosis
\Ensure Binary prediction: in-hospital death within 30 days

\State \textbf{Step 1: Patient Selection}
\State Identify patients using ICD-9 codes for traumatic brain injury (80000–85419)
\State Exclude patients aged $<65$, ICU stay $<24h$, missing vital signs (GCS, MAP, HR, SpO$_2$, Temp)
\State Retain only the first ICU admission per patient

\State \textbf{Step 2: Data Preprocessing}
\ForAll{numerical variables}
    \State Aggregate min, max, and mean over first 24 hours
\EndFor
\ForAll{categorical variables}
    \State Apply label encoding
\EndFor
\State Impute missing values using Random Forest-based imputation
\State Normalize all continuous variables using z-score standardization

\State \textbf{Step 3: Feature Selection}
\State Remove features with $>80\%$ missingness
\State Apply Random Forest importance (Gini-based) via RFECV
\State Retain 17 features from intersection of both methods

\State \textbf{Step 4: Model Development}
\State Perform stratified 70/30 train-test split
\ForAll{models $\in$ \{XGBoost, RF, AdaBoost, SVM, NB, LR\}}
    \State Tune hyperparameters via GridSearchCV (5-fold CV)
    \State Evaluate metrics: AUROC, accuracy, sensitivity, specificity, F1-score
\EndFor

\State \textbf{Step 5: Statistical Validation}
\State Compare training and test cohorts using t-test and chi-square test
\State Compute 95\% CI for AUROC via 2000 bootstrap replicates
\State Plot calibration curve and calculate Brier score

\State \textbf{Step 6: Model Interpretation}
\State Apply SHAP to final model to compute global and local feature importance
\State Perform ablation study: iteratively drop one feature and re-evaluate AUROC
\State Visualize SHAP summary and dependence plots
\end{algorithmic}
\end{algorithm}

\newpage
Patients were identified from the MIMIC-III database using standardized ICD-9 head trauma codes, with exclusion criteria applied to remove individuals under 65 years of age, ICU stays under 24 hours, and cases with missing vital signs. The final analytic cohort consisted of 667 patients.

Preprocessing steps included aggregation of numerical features across the first 24 hours (min, max, mean), unit harmonization, and z-score standardization. Missing values were imputed using Random Forest-based methods to preserve feature interdependencies.

From an initial 69 clinically relevant features informed by prior studies, we removed variables with more than 80\% missingness and applied a hybrid feature selection strategy using Random Forest importance (via Gini impurity). This process yielded 17 final features spanning demographics, comorbidities, vital signs, and functional scores.

Six machine learning models were trained and tuned using 5-fold cross-validation with stratified sampling. XGBoost was selected as the final classifier due to its superior performance on AUROC, accuracy, and interpretability. Hyperparameters were optimized to jointly maximize discrimination and calibration.

Statistical validation procedures ensured model robustness. Cohort comparability was confirmed using independent t-tests. Feature-level contributions were assessed through ablation studies, and interpretability was achieved via SHAP, revealing clinically plausible predictors such as GCS, heart rate, and oxygen saturation.

This comprehensive and reproducible framework supports the development of clinically interpretable and generalizable models for ICU risk stratification in elderly TBI patients.

\subsection*{Patient Selection}
Patients diagnosed with TBI were retrospectively identified from the publicly available MIMIC-III database \cite{johnson2016mimic} using International Classification of Diseases, Ninth Revision (ICD-9) codes indicative of head trauma. Specifically, codes ranging from 80000–80199, 80300–80499, and 85000–85419 were used, resulting in an initial pool of 3,025 unique ICU admissions. Structured Query Language (SQL) queries were executed in a PostgreSQL environment to ensure reproducibility and adherence to cohort selection logic.

A series of exclusion criteria were applied to enhance the homogeneity and clinical relevance of the final study cohort. First, 1,635 patients younger than 65 years were excluded to specifically target the geriatric TBI population, yielding 1,390 older adult ICU admissions. Second, patients lacking GCS assessments within the first 24 hours of ICU admission were excluded. GCS values were extracted from the CHARTEVENTS table using ITEMIDs 223900 (MetaVision) and 198 (CareVue). After applying this criterion, 30 additional patients were excluded.

Third, completeness of core vital signs was required for inclusion. Specifically, patients were excluded if they lacked any of the following physiological measurements during the first 24 hours of ICU stay: heart rate, non-invasive mean arterial pressure (MAP), respiratory rate, peripheral oxygen saturation (SpO$_2$), and body temperature (recorded in Fahrenheit). Following this step, 693 patients were removed due to missing data, resulting in a final analytic cohort of 667 geriatric TBI patients.

For all included patients, clinical features encompassing demographics, comorbidities, vital signs, and laboratory data were extracted from the first 24 hours of ICU admission. This approach ensured that all model inputs reflected early physiological status, consistent with the goal of early risk prediction for in-hospital mortality.

\begin{figure}[H]
\centering
\includegraphics[width=0.85\linewidth]{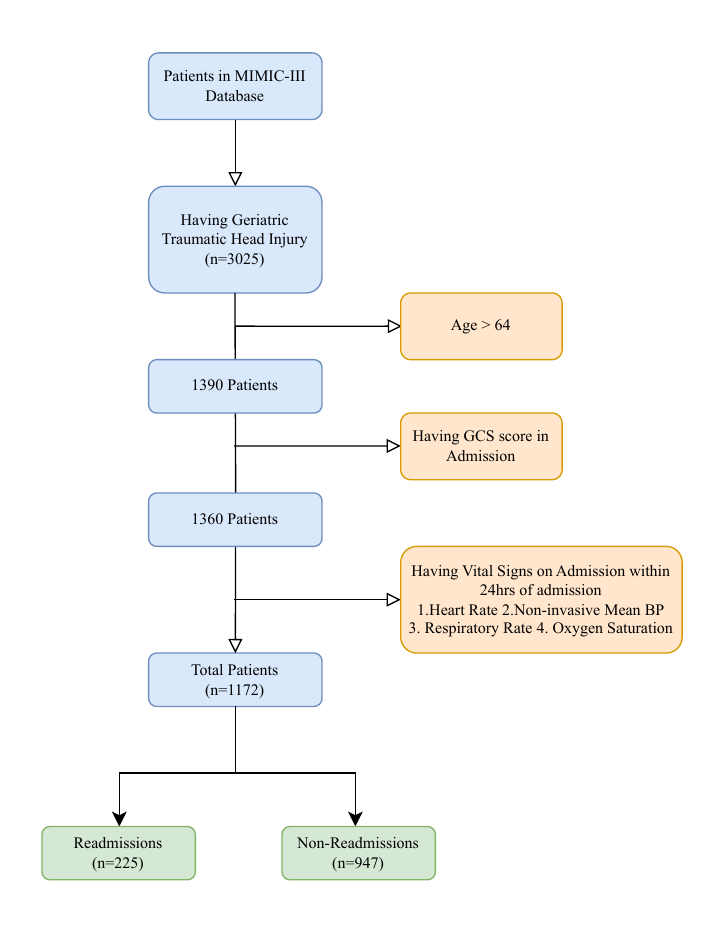} 
\caption{\textbf{Flowchart Illustrating the Patient Selection Process from the MIMIC-III Database}}
\label{fig:patient_selection}
\end{figure}

 The complete patient extraction workflow is illustrated in Figure~\ref{fig:patient_selection}.

\subsection*{Data Preprocessing}

A rigorous, multi-stage data preprocessing strategy was implemented to ensure data integrity, internal consistency, and analytical validity prior to model development. This step was crucial for addressing common challenges in clinical datasets, including missingness, heterogeneous measurement units, and feature scale imbalances—each of which can significantly compromise model accuracy, generalizability, and interpretability if left unaddressed \cite{Si2025.03.14.25324005, fan2025lightgbm}.

To address missing data, we employed a Random Forest-based imputation method. Unlike conventional univariate imputation techniques such as mean or median substitution, Random Forests model each variable with missing values as a function of all other available features, thereby preserving nonlinear inter-variable relationships and enhancing the clinical realism of the imputed dataset. This method is particularly advantageous for ICU data, where multivariate dependencies among vital signs and laboratory variables are critical for reliable risk modeling. Additionally, its compatibility with both categorical and continuous data types enabled comprehensive treatment of the dataset's heterogeneous feature space.

Post-imputation, unit consistency was enforced across the dataset. For example, temperature values recorded in both Fahrenheit and Celsius were converted to a common scale. Furthermore, variables such as patient age and ICU length of stay were calculated using precise timestamp differentials to ensure temporal accuracy and data reproducibility.

To capture clinically informative variability during the early ICU period, key physiological variables—such as heart rate, glucose levels, and blood pressure—were transformed into derived features using descriptive statistics over the first 24 hours of ICU admission:

\begin{equation}
x_{\text{min}} = \min_{t \in [1, T]} x_t, \quad
x_{\text{max}} = \max_{t \in [1, T]} x_t, \quad
x_{\text{mean}} = \frac{1}{T} \sum_{t=1}^{T} x_t
\label{eq:temporal_stats}
\end{equation}

This aggregation strategy enabled the model to capture both static and dynamic aspects of patient physiology—elements known to be highly predictive of acute deterioration in critical care settings.

To mitigate the risk of scale-driven model bias, all continuous variables were normalized using z-score standardization:

\begin{equation}
z_i = \frac{x_i - \mu}{\sigma}
\label{eq:zscore}
\end{equation}

where \( \mu \) and \( \sigma \) denote the mean and standard deviation of each variable, computed exclusively from the training dataset. This transformation ensured that all features contributed equitably to gradient-based learning algorithms, especially tree-based and neural network models that are sensitive to variable magnitude.

Finally, patients who did not meet the predefined clinical inclusion criteria—such as age under 65 years, ICU stay duration less than 24 hours, or missing essential physiological measurements—were excluded to ensure cohort homogeneity. Only the first ICU stay was retained per patient to eliminate duplication bias and avoid skewing outcome distributions.

Collectively, these preprocessing procedures yielded a clean, harmonized, and temporally enriched dataset with preserved clinical structure and minimized noise. This high-quality analytical input formed the foundation for robust, interpretable, and generalizable modeling of 30-day mortality among geriatric TBI patients.

\subsection*{Feature Selection}

\begin{figure}[htbp]
    \centering
    \includegraphics[width=1\linewidth]{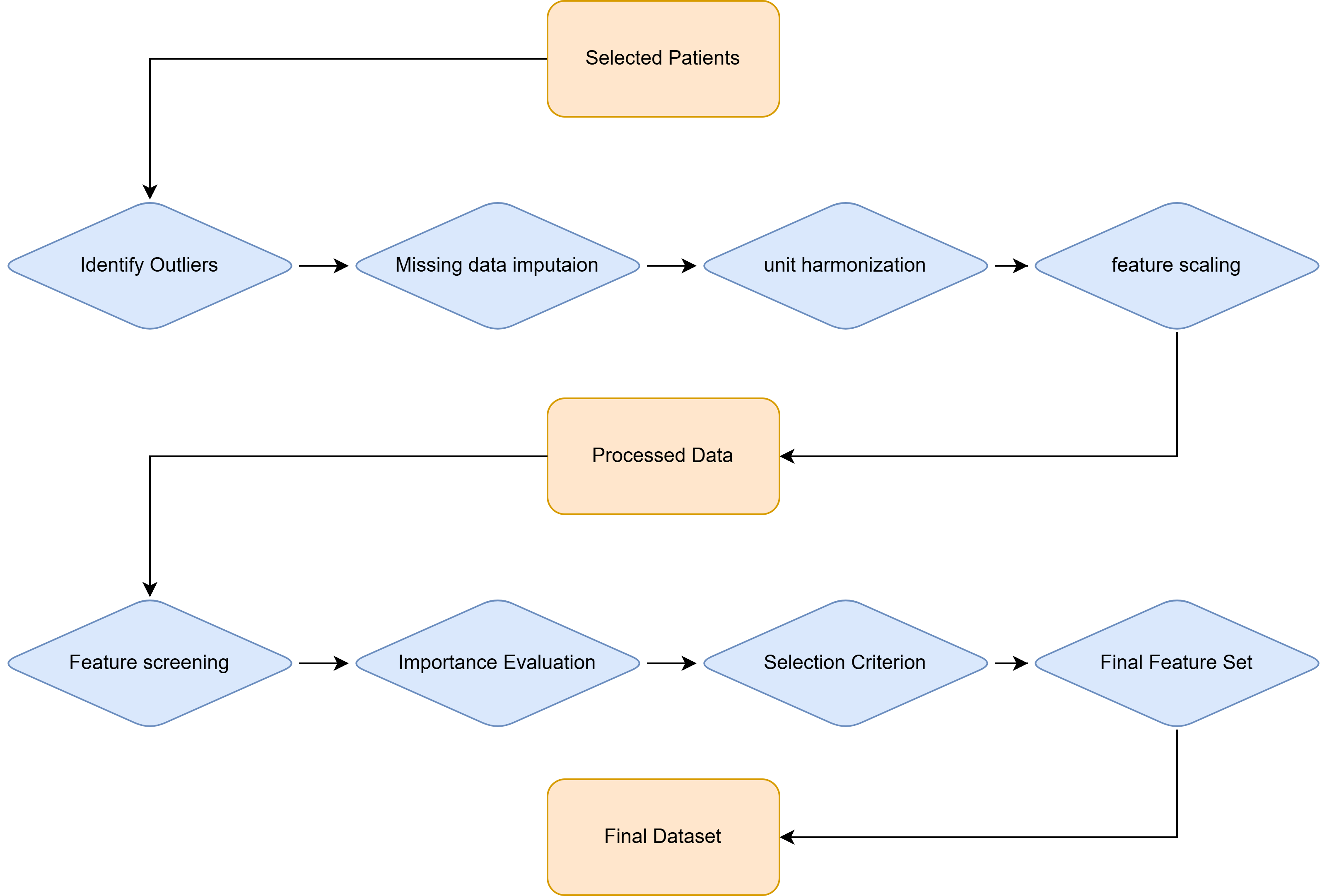}
    \caption{\textbf{Flowchart Illustrating the Full Data Preprocessing and Feature Selection Pipeline}}
    \label{feature_selection.png}
\end{figure}

A supervised feature selection strategy was adopted to identify the most informative variables for predicting 30-day mortality in geriatric TBI patients. The full pipeline, including variable filtering, importance ranking, and final selection, is illustrated in Figure~\ref{feature_selection.png}.

To establish a clinically interpretable and statistically robust input space for model development, we began by compiling a set of 69 candidate features. These variables were selected based on prior literature in TBI prognosis modeling and consultations with clinical experts specializing in neurocritical care \cite{wang2023brainsci,sun2025optimizingurbanmobilitycomplex}. The initial features were systematically categorized into three primary domains: 

\begin{itemize}
\item \textbf{Demographics and Administrative Indicators}, such as patient age, marital status, ethnicity, and type of insurance, were included to account for population-level heterogeneity and care access disparities.
\item \textbf{Comorbidities}, including cardiovascular disease, diabetes, chronic respiratory illness, and dementia, were incorporated to reflect pre-existing conditions known to influence short-term outcomes after TBI.
\item \textbf{Clinical Measurements}, which encompassed vital signs (heart rate, respiratory rate, temperature, oxygen saturation), neurological assessment (GCS score), and laboratory values (e.g., prothrombin time), captured the acute physiological status within the first 24 hours of ICU admission.
\end{itemize}

To ensure data quality, we first excluded variables with more than 80\% missingness, which resulted in the removal of features such as arterial lactate, AST/ALT levels, and non-invasive systolic blood pressure—variables that, although clinically relevant, lacked sufficient coverage for robust modeling.

Subsequently, we employed a model-driven feature selection strategy leveraging Random Forest–based importance scores. Random Forest was chosen over other ensemble classifiers (e.g., XGBoost, AdaBoost) due to its demonstrated stability across variable scales and its ability to handle mixed data types with minimal preprocessing. Feature importance was quantified using the Gini impurity reduction criterion aggregated across all trees in the forest, defined as:

\begin{equation}
I(x_i) = \sum_{t \in T} \frac{p(t) \cdot \Delta i(t)}{f(t)}
\label{eq:gini}
\end{equation}

where \( I(x_i) \) denotes the importance score of feature \( x_i \), \( T \) is the set of all nodes in all trees of the random forest, \( p(t) \) is the proportion of samples reaching node \( t \), \( \Delta i(t) \) is the decrease in Gini impurity at node \( t \), and \( f(t) \) is the frequency with which feature \( x_i \) appears in tree splits.

This process enabled the identification of variables that consistently contributed to accurate classification while discarding those with marginal or redundant information. For instance, several comorbidities (e.g., metastatic cancer, paraplegia) and redundant physiological indicators (e.g., both maximum and mean temperature) were excluded due to limited predictive value or multicollinearity.

In total, 9 features were selected for model training. These final predictors strike a balance between parsimony and clinical representativeness, covering essential dimensions such as neurological status (e.g., \textit{GCS\_SCORE}), physiological stability (e.g., \textit{PT}, \textit{Oxygen Saturation}), and baseline demographics (e.g., \textit{AGE}, \textit{ETHNICITY\_E}). This feature set served as the foundation for all downstream modeling and interpretability analyses.

\begin{table}[H]
\noindent
\caption{\textbf{Final 9 Features Used for 30-day Mortality Prediction in Geriatric TBI Patients.}}
\label{tab:final_features}
\small
\renewcommand{\arraystretch}{1.2}
\rowcolors{2}{white}{white}
\begin{tabularx}{\textwidth}{l|>{\raggedright\arraybackslash}X}
\hline
\rowcolor[HTML]{D9EAD3}
\textbf{Category} & \textbf{Features} \\
\hline
Demographics & Age (\textit{AGE}), Marital Status (\textit{MARITAL\_STATUS\_E}), Ethnicity (\textit{ETHNICITY\_E}) \\
\hline
Administrative & Emergency Department Length of Stay (\textit{ED\_LENGTH\_OF\_STAY}) \\
\hline
Clinical Measurements & Glasgow Coma Scale Score (\textit{GCS\_SCORE}), Temperature (\textit{Temperature\_Combined}), Prothrombin Time (\textit{PT}), Heart Rate, Oxygen Saturation \\
\hline
\end{tabularx}
\end{table}

\subsection*{Model Development and Evaluation}

To predict 30-day mortality among elderly ICU patients with TBI, we implemented a supervised learning framework using the 9 selected features. The dataset was randomly partitioned into a training set (70\%) and a holdout test set (30\%) using stratified sampling to preserve outcome distributions.

\begin{figure}[htbp]
\centering
\includegraphics[width=1\linewidth]{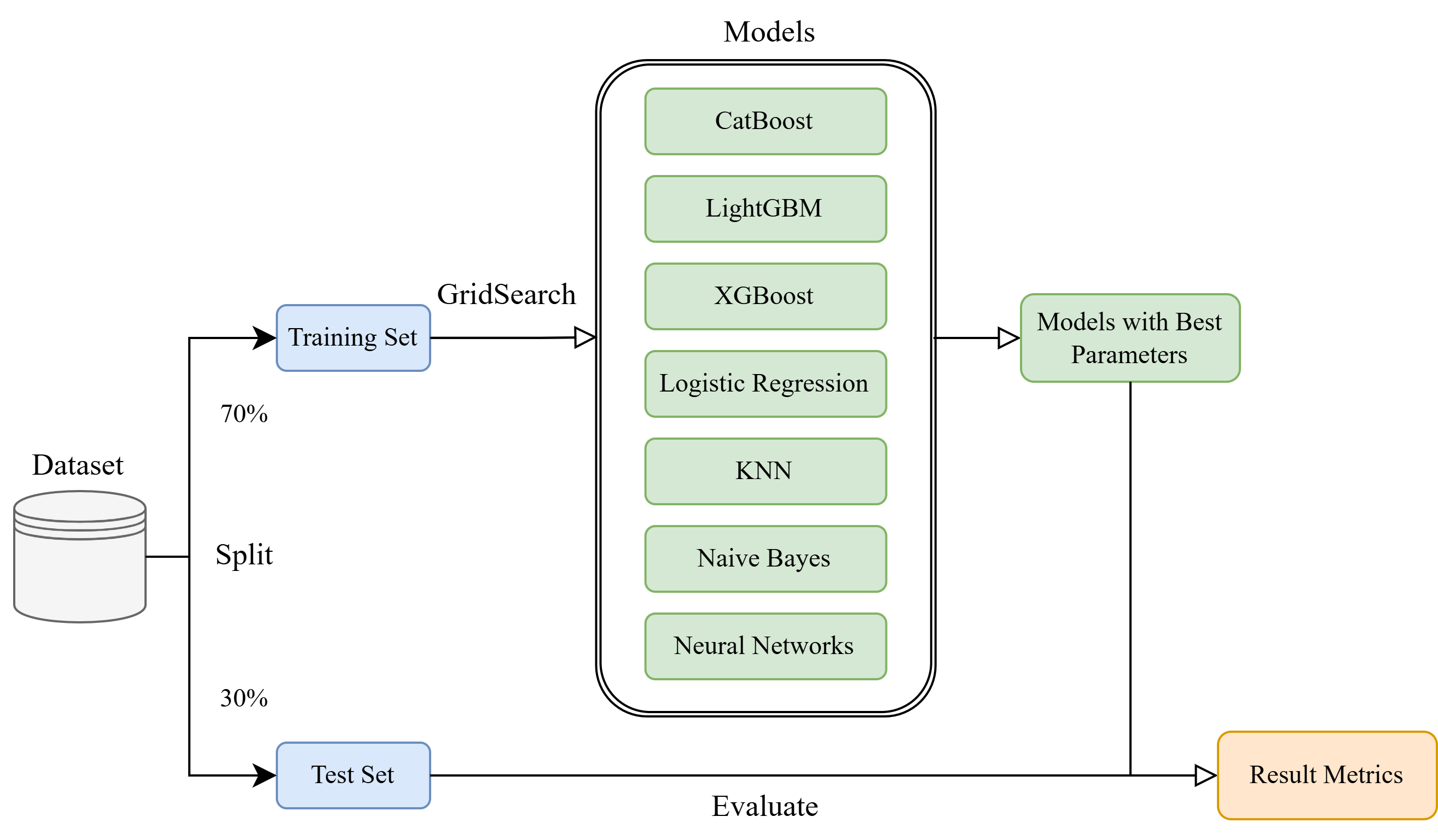}
\caption{\textbf{Flowchart of Multi-model Development and Evaluation.}}
\label{fig:model}
\end{figure}

We developed and compared seven machine learning classifiers with complementary algorithmic principles and empirical advantages: CatBoost, LightGBM, XGBoost, Logistic Regression (LR), K-Nearest Neighbors (KNN), Gaussian Naïve Bayes (NB), and a fully connected Neural Network (NeuralNet). Each model was trained using grid search with stratified 5-fold cross-validation to ensure robust hyperparameter selection. Final evaluations were conducted exclusively on the unseen test set to assess generalization performance.Figure~\ref{fig:model} provides a schematic overview of the modeling pipeline.

CatBoost was chosen for its ability to effectively handle categorical features and small datasets through ordered boosting and symmetric tree structures. Its unique approach to encoding and its inherent resistance to overfitting made it particularly well-suited for clinical scenarios where variable interdependence and missingness are common. Similarly, LightGBM offered a highly efficient gradient boosting solution, employing histogram-based binning and leaf-wise tree growth to accelerate training without sacrificing performance. This made it advantageous for scenarios involving high-dimensional or partially sparse input vectors. XGBoost, with its regularized learning objective and second-order optimization, served as a robust and well-established benchmark, particularly adept at capturing subtle nonlinearities and interactions across features. For all boosting models, extensive grid search was conducted to optimize key hyperparameters, including learning rate, tree depth, regularization penalties, and feature sampling ratios, thereby ensuring stable convergence and generalizability.

To provide an interpretable linear baseline, we incorporated Logistic Regression, which retains clinical utility due to its direct interpretability and well-understood statistical foundations. In this model, both L1 (Lasso) and L2 (Ridge) penalties were evaluated to manage multicollinearity and enhance model sparsity, with the inverse regularization strength (\texttt{C}) tuned to balance complexity and fit. K-Nearest Neighbors was included to assess the impact of local neighborhood structures on classification accuracy. As a non-parametric and instance-based learner, KNN is particularly sensitive to the spatial distribution of samples and thus provides insight into the geometric separability of the feature space. Hyperparameters such as the number of neighbors, distance metric, and weighting schemes were systematically tuned.

Gaussian Naïve Bayes, despite its strong independence assumptions, was incorporated as a lightweight probabilistic benchmark due to its resilience in high-dimensional settings and computational efficiency. Its generative nature and closed-form estimators allowed for rapid evaluation, offering a contrast to more complex discriminative models. Finally, a shallow feedforward Neural Network was employed to evaluate the expressive capacity of flexible, nonlinear function approximators. The network, consisting of a single hidden layer with ReLU activation and a sigmoid output, was trained using the Adam optimizer. Hyperparameters including neuron count, learning rate, batch size, and dropout rate were optimized to mitigate overfitting while enabling the model to capture latent interactions.

All models were trained using stratified 5-fold cross-validation with grid search optimization on the training set. The independent test set was withheld from all training and validation procedures and used exclusively for final evaluation, ensuring an unbiased estimate of each model's generalization performance.

Model performance was evaluated using the AUROC on the test set. Robustness was assessed by computing 95\% confidence intervals via 2,000 bootstrap replicates. Letting \( \hat{y}_i \) denote predicted probabilities and \( y_i \) the ground-truth labels, models were trained to minimize a regularized loss function:

\begin{equation}
\mathcal{L}(\theta) = \sum_{i=1}^{n} \ell(y_i, \hat{y}_i) + \Omega(f),
\label{eq:loss_function}
\end{equation}

where \( \ell \) is the binary cross-entropy loss and \( \Omega(f) \) penalizes model complexity.

This structured modeling pipeline ensured a fair comparison of diverse classifiers, controlled for overfitting, and yielded a robust and interpretable predictive model capable of stratifying 30-day mortality risk in geriatric TBI patients.

\subsection*{Statistical Evaluation and Interpretability Framework}

To ensure the scientific validity, robustness, and clinical applicability of the proposed machine learning model for 30-day mortality risk prediction in patients with TBI, we implemented a structured three-part statistical evaluation strategy: (1) assessing the comparability of training and test cohorts via t-test, (2) quantifying feature contribution through ablation analysis, and (3) enhancing transparency via SHAP.

First, we evaluated cohort equivalence to confirm that the training and validation datasets were statistically comparable. Two-sided Student’s t-tests were conducted on continuous clinical variables such as age, vital signs, and laboratory results:

\begin{equation}
t = \frac{\bar{X}_1 - \bar{X}_2}{\sqrt{s_p^2 \left( \frac{1}{n_1} + \frac{1}{n_2} \right)}}
\label{eq:t_test}
\end{equation}

where \( \bar{X}_1 \) and \( \bar{X}_2 \) represent group means and \( s_p^2 \) is the pooled variance. Welch’s correction was applied when variance homogeneity was not met. This step ensured that model evaluation was not confounded by baseline differences between patient cohorts, thereby reinforcing the credibility of performance metrics.

Second, we performed an ablation analysis to assess the relative importance of individual features to the model’s predictive power. This involved iteratively removing each feature from the input set and measuring the change in model performance:

\begin{equation}
\Delta_i = \text{AUROC}(f_{\text{full}}) - \text{AUROC}(f_{-i})
\label{eq:ablation}
\end{equation}

where \( f_{\text{full}} \) and \( f_{-i} \) represent the models with and without feature \( x_i \), respectively. This approach provides an intuitive and quantifiable way to identify which features are most influential to prediction outcomes. Such insights support feature prioritization for clinicians and inform the design of simplified risk scores or targeted monitoring protocols.

Third, we employed SHAP to improve interpretability at both global and individual levels. SHAP computes additive contributions of each feature to the final model output for a given patient:

\begin{equation}
\phi_i = \sum_{S \subseteq N \setminus \{i\}} \frac{|S|!(|N|-|S|-1)!}{|N|!} \left[ f(S \cup \{i\}) - f(S) \right]
\label{eq:shapley}
\end{equation}

where \( N \) is the set of all input features. Unlike traditional importance rankings, SHAP allows real-time interpretation of model predictions for individual patients, offering clinicians actionable insight into which clinical variables contributed most to the estimated risk. This supports transparency, facilitates shared decision-making, and enhances trust in the model’s use as a bedside decision-support tool.

Together, these validation strategies not only verified the methodological soundness of the predictive framework but also ensured that the model’s outputs could be meaningfully interpreted and translated into clinical contexts.

\section*{Results}
\subsection*{Cohort Characteristics and Statistical Comparison}

The data set analyzed in this study includes hospitalized patients diagnosed with TBI. Patients were randomly assigned into a training cohort (70\%, n=820) and a test cohort (30\%, n=352) to develop and validate machine learning models for predicting 30-day mortality. Separately, patients were grouped based on survival outcomes, with 947 patients surviving (Group 1) and 225 patients not surviving (Group 0) within 30 days.

The cohort split was designed to ensure that the machine learning models could generalize effectively to unseen patients. To validate this, it is critical that the training and test sets maintain similar clinical profiles. Table~\ref{tab:cohort comparison results} summarizes statistical comparisons in eight key clinical characteristics, presenting means, standard deviations, and associated p-values. A significance threshold of 0.05 was adopted. Most of the features, including the GCS score, length of stay in the ICU, temperature, heart rate, and oxygen saturation, did not show statistically significant differences between training and test sets. This suggests that the two cohorts are well matched, reducing the likelihood of sampling bias and supporting the external validity of the model.

Furthermore, Table~\ref{tab:cohort comparison results 1} presents comparisons between patients who survived versus those who did not. Several features exhibited statistically significant differences. In particular, patients who survived had significantly higher GCS scores (11.94 vs. 7.20, p < 0.001) and higher oxygen saturation levels (98.37\% vs. 97.68\%, p = 0.0448), reflecting better neurological and respiratory status. In contrast, patients who did not survive tended to have a prolonged prothrombin time (14.51 vs 13.83, p = 0.0176), which may indicate underlying coagulopathy or systemic injury. Other variables, including heart rate and temperature, showed marginal differences between the groups.

These findings have two critical implications. First, the clinical similarity between the training and test cohorts ensures that the model is not overfitted to peculiarities within the training data, thereby improving its generalizability. Second, identifying specific characteristics such as the GCS score, oxygen saturation, and prothrombin time that differ significantly between survival groups improves the clinical interpretability of the model and aligns with the established medical understanding of the severity and prognosis of TBI.

By combining robust statistical validation with meaningful clinical interpretation, this study supports the development of machine learning models that are both technically reliable and clinically applicable. Structured comparison of feature distributions provides transparency and increases confidence among clinicians and researchers, particularly those less familiar with machine learning methodologies, that the models are grounded in real-world patient characteristics and clinical reasoning.

\begin{table}[H]
\noindent
\caption{\textbf{T-test Comparison of Feature Distributions between Training and Test Sets.}}
\label{tab4}
\small
\renewcommand{\arraystretch}{1.2}
\rowcolors{2}{white}{white}
\begin{tabularx}{\textwidth}{>{\raggedright\arraybackslash}X|X|X|X}
\hline
\rowcolor[HTML]{D9EAD3}
{\bf Feature} & {\bf Training Set} & {\bf Test Set} & {\bf P-value} \\ \hline
GCS\_SCORE & 10.80 (4.28) & 11.05 (4.32) & 0.489 \\ \hline
Temperature & 97.88 (4.90) & 97.76 (2.01) & 0.647 \\ \hline
ED\_LOS & 5.31 (4.89) & 4.84 (3.43) & 0.156 \\ \hline
PT & 13.99 (2.78) & 15.04 (10.17) & 0.153 \\ \hline
Heart Rate & 81.22 (16.96) & 80.30 (15.99) & 0.506 \\ \hline
AGE & 111.46 (78.97) & 103.96 (71.14) & 0.228 \\ \hline
Oxygen Saturation & 97.85 (3.14) & 97.92 (2.61) & 0.739 \\ \hline
MARITAL\_STATUS & 0.23 (0.06) & 0.22 (0.05) & 0.262\\ \hline
ETHNICITY\_E & 0.25 (0.07) & 0.24 (0.07) & 0.048 \\ \hline
\end{tabularx}
\begin{flushleft}
Table notes: The table summarizes differences between the training and test cohorts across multiple clinical variables. Mean and standard deviation (SD) are reported. P-values are computed using appropriate statistical tests (e.g., t-test) with a significance threshold of 0.05.
\end{flushleft}
\label{tab:cohort comparison results}
\end{table}

\begin{table}[H]
\noindent
\caption{\textbf{T-test Comparison of Feature Distributions between Survival and Non-Survival Sets.}}
\label{tab4}
\small
\renewcommand{\arraystretch}{1.2}
\rowcolors{2}{white}{white}
\begin{tabularx}{\textwidth}{>{\raggedright\arraybackslash}X|X|X|X}
\hline
\rowcolor[HTML]{D9EAD3}
{\bf Feature} & {\bf Non-Survival} & {\bf Survival} & {\bf P-value} \\ \hline
GCS\_SCORE & 11.94 (3.68) & 7.20 (4.02) & 0.000\\ \hline
Temperature & 98.04 (5.45) & 97.36 (2.33) & 0.063 \\ \hline
ED\_LOS & 5.62 (5.23) & 4.32 (3.44) & 0.003 \\ \hline
PT & 13.83 (2.82) & 14.51 (2.58) & 0.018 \\ \hline
Heart Rate & 80.37 (16.11) & 83.89 (19.25) & 0.080 \\ \hline
AGE & 109.26 (76.87) & 118.44 (85.25) & 0.311 \\ \hline
Oxygen Saturation & 97.68 (3.14) & 98.37 (3.13) & 0.045 \\ \hline
MARITAL\_STATUS & 0.22 (0.05) & 0.24 (0.08) & 0.119 \\ \hline
ETHNICITY\_E & 0.24 (0.07) & 0.27 (0.08) & 0.000 \\ \hline
\end{tabularx}
\begin{flushleft}
Table notes: This table compares patients survive or dead within 30 days. Differences in mean values of key clinical variables are shown along with p-values. Statistical significance set at 0.05 threshold.
\end{flushleft}
\label{tab:cohort comparison results 1}
\end{table}

\subsection*{Ablation Study and Feature Contribution Analysis}

\begin{figure}[H]
\begin{adjustwidth}{-2.25in}{0in}
    \centering
    \includegraphics[width=0.95\linewidth]{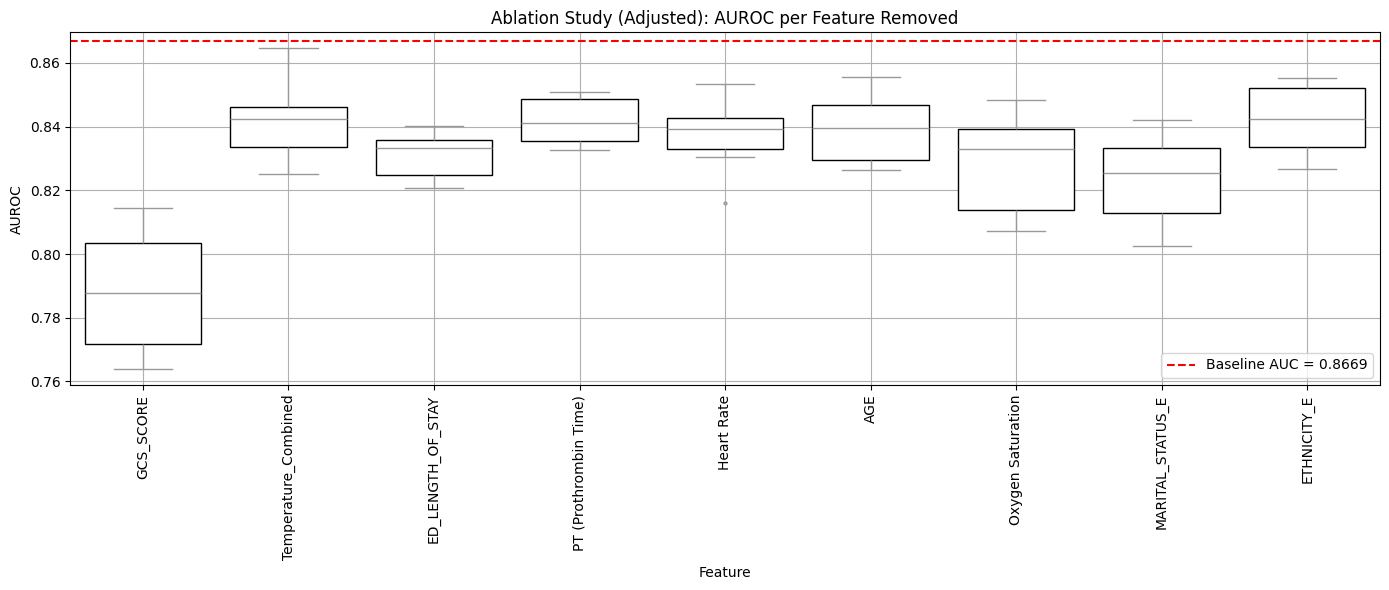}
    \caption{\textbf{Impact of Feature Removal on CatBoost Model Performance.}}
    \label{fig:ablation analysis}
\end{adjustwidth}
\end{figure}

To further assess the robustness and clinical relevance of our predictive model for 30-day mortality in TBI patients, we performed an ablation analysis, as illustrated in Fig.~\ref{fig:ablation analysis}. In this experiment, each feature was systematically removed from the input set, and the model was retrained on the modified dataset. Performance was evaluated over ten bootstrapped iterations to ensure statistical stability. The baseline model was a fully trained CatBoost classifier, achieving an AUROC of 0.8669 on the test set when using all available features.

Figure~\ref{fig:ablation analysis} presents the distribution of the AUROC scores after the removal of each clinical characteristic. Each boxplot represents the AUROC variability across the ten resampling rounds, with the median indicated by the central line and the interquartile range captured by the bounds of the box. The whiskers extend 1.5 times the interquartile range, and a red dashed line marks the baseline AUROC achieved with all included features, serving as a comparative reference.

The exclusion of key features, particularly the GCS score, resulted in the most substantial decline in AUROC, underscoring its critical role in model performance. Other variables such as the duration of stay in the ED, the time to prothrombin, and oxygen saturation also demonstrated noticeable impacts when removed, suggesting their significant contribution to the prediction of mortality in patients with TBI. Even features like marital status and ethnicity, which might appear less significant in univariate analyses, influenced performance in the full multivariate context, emphasizing the importance of complex feature interactions.

In particular, the general trend that removal of any single feature led to a decrease in AUROC supports the notion that the model does not overly depend on a single dominant predictor. Instead, it integrates a broad range of physiological and demographic inputs, enhancing its generalizability and robustness across diverse clinical populations.

From a clinical perspective, the importance of characteristics such as the GCS score, oxygen saturation, and prothrombin time is consistent with established understandings of neurological function, respiratory status, and coagulopathy in the prognosis of TBI. The model’s reliance on these clinically intuitive variables strengthens its interpretability and potential utility for bedside decision-making.

In summary, the ablation study validates the resilience of the predictive model and highlights the complementary contributions of multiple clinical factors. Rather than relying on isolated variables, model performance is derived from the collective synthesis of diverse clinical signals, strengthening both statistical rigor and clinical credibility.

\subsection*{Model Performance Evaluation and Comparative Analysis}

\begin{figure}[H]
\centering
\includegraphics[width=0.95\linewidth]{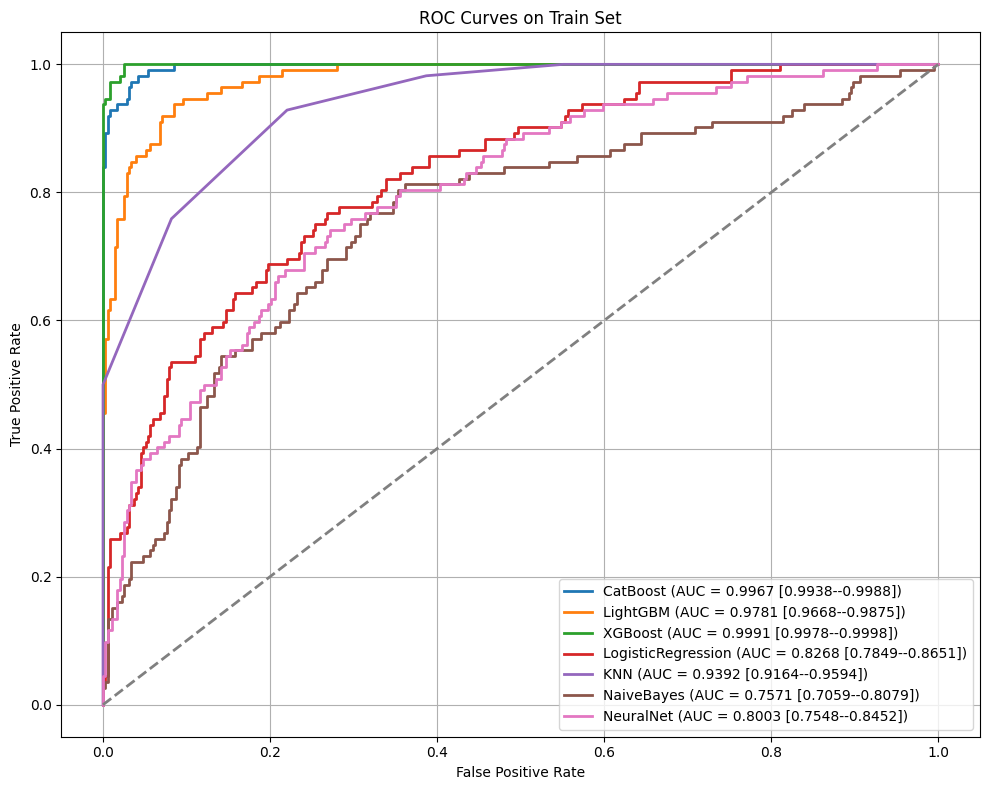}
\caption{\textbf{AUROC Curves for Model Performance in the Training Set.}}
\label{fig:roc_train}
\end{figure}

\begin{figure}[H]
\centering
\includegraphics[width=0.95\linewidth]{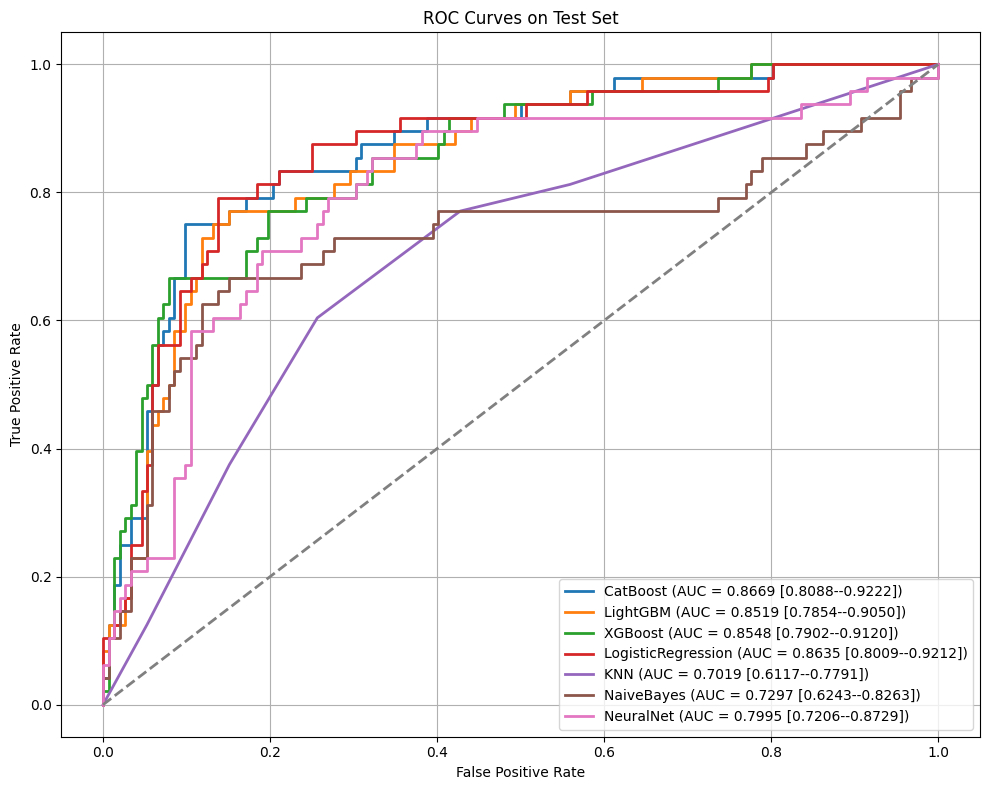}
\caption{\textbf{AUROC Curves for Model Performance in the Test Set.}}
\label{fig:roc_test}
\end{figure}

To comprehensively assess the performance and generalizability of various machine learning models in predicting 30-day mortality among patients with TBI, we systematically compared seven classifiers across both the training and test sets. Figure~\ref{fig:roc_train} and Figure~\ref{fig:roc_test} display the ROC curves for the training and test cohorts, respectively, while Table~\ref{tab: Results of the Training Set} and Table~\ref{tab: Results of the Test Set} present detailed performance metrics, including AUROC, accuracy, sensitivity, specificity, PPV, NPV, and F1 score with corresponding 95\% confidence intervals.

On the training set, ensemble models such as CatBoost, XGBoost, and LightGBM exhibited superior learning ability. XGBoost achieved an AUROC of 0.9991, followed closely by CatBoost at 0.9967 and LightGBM at 0.9781. These models demonstrated excellent sensitivity and specificity, suggesting a strong capacity to model complex, nonlinear interactions in high-dimensional clinical datasets. In contrast, models like Logistic Regression, NaiveBayes, and NeuralNet showed relatively lower performance, underscoring potential limitations in capturing the intricate physiological patterns seen in TBI patients.

However, as performance on the training set alone can be misleading due to potential overfitting, evaluation on the independent test set was critical. CatBoost maintained the highest AUROC at 0.8669 (95\% CI: 0.8088–0.9222), along with a balanced sensitivity (0.7517) and specificity (0.8879), indicating robust predictive power. LightGBM and XGBoost also performed competitively, achieving AUROCs of 0.8519 and 0.8548, respectively. Logistic Regression performed surprisingly well compared to other simpler models, with an AUROC of 0.8635. In contrast, KNN, NaiveBayes, and NeuralNet demonstrated diminished generalization, especially in PPV (all <0.45), suggesting limited clinical applicability.

The strong generalization ability of CatBoost, LightGBM, and XGBoost is largely attributable to their tree-based boosting architectures, which naturally model nonlinearities and feature interactions without relying on strict parametric assumptions. These methods inherently accommodate the heterogeneity and missingness often encountered in ICU datasets, making them particularly well-suited for real-world clinical data.

Among these, CatBoost exhibited particular advantages. Its ordered boosting mechanism and optimal handling of categorical variables likely contributed to more stable learning, even in the presence of class imbalance—an important consideration given the smaller proportion of mortality events in our cohort. Compared to XGBoost and LightGBM, CatBoost also demonstrated slightly higher NPV (0.9182) and PPV (0.6793), critical indicators for minimizing false negatives and ensuring efficient allocation of clinical resources.

From a clinical standpoint, CatBoost’s ability to maintain high sensitivity and NPV is especially valuable in mortality prediction, where the cost of missing a high-risk patient can be substantial. The balanced performance across sensitivity and specificity, combined with clinically interpretable feature contributions demonstrated in prior analyses, enhances the model’s potential utility for bedside deployment.

In summary, CatBoost emerged as the most effective and clinically interpretable model for predicting 30-day mortality in TBI patients. Its balanced accuracy, strong minor class handling, and compatibility with the clinical data structure underscore its suitability for integration into real-world clinical decision support systems.

\begin{table}[H]
\small
\renewcommand{\arraystretch}{1.2}
\begin{adjustwidth}{-2.25in}{0in}
\centering
\caption{\textbf{Performance Comparison of Different Models in the Training Set.}}
\begin{tabular}{l|l|l|l|l|l|l|l}
\hline
\rowcolor[HTML]{D9EAD3}
\textbf{Model} & \textbf{AUROC (95\% CI)} & \textbf{Accuracy} & \textbf{F1-score} & \textbf{Sensitivity} & \textbf{Specificity} & \textbf{PPV} & \textbf{NPV} \\ \hline
\rowcolor[HTML]{FDE9D9}
CatBoost & 0.997 (0.994--0.999) & 0.974 & 0.945 & 0.929 & 0.988 & 0.964 & 0.978 \\ \hline
LightGBM & 0.978 (0.967--0.988) & 0.927 & 0.849 & 0.855 & 0.949 & 0.842 & 0.955 \\ \hline
XGBoost & 0.999 (0.998--1.000) & 0.985 & 0.968 & 0.965 & 0.992 & 0.973 & 0.989 \\ \hline
LogisticRegression & 0.827 (0.785--0.865) & 0.746 & 0.582 & 0.749 & 0.745 & 0.481 & 0.904 \\ \hline
KNN & 0.939 (0.916--0.959) & 0.814 & 0.708 & 0.928 & 0.780 & 0.572 & 0.972 \\ \hline
NaiveBayes & 0.757 (0.706--0.808) & 0.643 & 0.522 & 0.811 & 0.590 & 0.384 & 0.909 \\ \hline
NeuralNet & 0.800 (0.755--0.845) & 0.683 & 0.542 & 0.787 & 0.650 & 0.414 & 0.905 \\ \hline

\end{tabular}
\label{tab: Results of the Training Set}
\end{adjustwidth}

\end{table}

\begin{table}[H]
\small
\renewcommand{\arraystretch}{1.2}
\begin{adjustwidth}{-2.25in}{0in}
\centering
\caption{\textbf{Performance Comparison of Different Models in the Test Set.}}
\begin{tabular}{l|l|l|l|l|l|l|l}
\hline
\rowcolor[HTML]{D9EAD3}
\textbf{Model} & \textbf{AUROC (95\% CI)} & \textbf{Accuracy} & \textbf{F1-score} & \textbf{Sensitivity} & \textbf{Specificity} & \textbf{PPV} & \textbf{NPV} \\ \hline
\rowcolor[HTML]{FDE9D9}
\textbf{CatBoost} & \textbf{0.867 (0.809--0.922)} & 0.855 & 0.710 & 0.752 & 0.888 & 0.679 & 0.918 \\ \hline
LightGBM & 0.852 (0.785--0.905) & 0.831 & 0.678 & 0.752 & 0.855 & 0.618 & 0.914 \\ \hline
XGBoost & 0.855 (0.790--0.912) & 0.830 & 0.653 & 0.663 & 0.881 & 0.637 & 0.893 \\ \hline
LogisticRegression & 0.864 (0.801--0.921) & 0.812 & 0.668 & 0.814 & 0.808 & 0.577 & 0.931 \\ \hline
KNN & 0.702 (0.612--0.779) & 0.711 & 0.499 & 0.604 & 0.743 & 0.428 & 0.855 \\ \hline
NaiveBayes & 0.730 (0.624--0.826) & 0.674 & 0.515 & 0.731 & 0.657 & 0.401 & 0.885 \\ \hline
NeuralNet & 0.800 (0.721--0.873) & 0.709 & 0.587 & 0.855 & 0.662 & 0.445 & 0.935 \\ \hline

\end{tabular}
\label{tab: Results of the Test Set}
\end{adjustwidth}
\end{table}

\subsection*{SHAP Analysis and Clinical Interpretability}

\begin{figure}[H]
    \centering
    \includegraphics[width=0.85\linewidth]{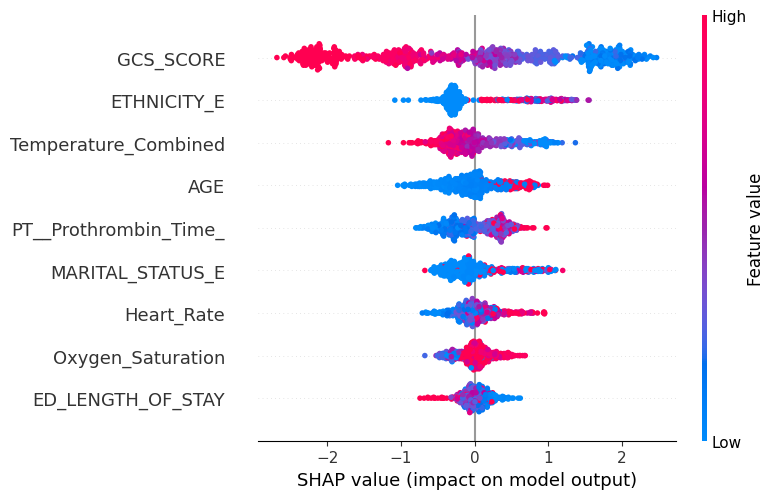}
    \caption{\textbf{SHAP Summary Plot Showing Feature Value Distributions and Their Impact on Model Output.}}
    \label{fig:shap_summary}
\end{figure}

To enhance the interpretability of the CatBoost model and ensure its alignment with clinical reasoning, SHAP values were employed to quantify the marginal contribution of each predictor to the model's output. The SHAP summary plot in Fig.~\ref{fig:shap_summary} displays the distribution of SHAP values for each variable in the prediction of 30-day mortality among patients with TBI.

Each row in the plot represents a feature, and individual points indicate SHAP values for each patient. The horizontal axis denotes the SHAP value - how much a feature shifts the prediction toward mortality or survival - while the color reflects the original feature value, ranging from low (blue) to high (red). Features positioned further to the right with high SHAP values exert a stronger influence in increasing the predicted mortality risk.

The GCS score emerged as the most influential predictor in the model. Low GCS values (blue) strongly shifted the prediction toward increased mortality, in line with the clinical understanding that impaired neurological status is a hallmark of poor prognosis in TBI. ETHNICITY\_E and PT (Prothrombin Time) followed, suggesting that coagulopathy and demographic-related disparities may significantly affect outcomes.

Temperature and age also played an important role. Lower temperatures and older age were associated with higher predicted mortality, which could reflect impaired thermoregulation and physiological reserve in critically ill patients. MARITAL\_STATUS\_E demonstrated modest but consistent effects, potentially capturing psychosocial factors that influence recovery and survival trajectories.

Additional contributors such as heart rate, oxygen saturation, and ED length of stay exhibited secondary influences. In particular, decreased oxygen saturation (blue points on the right) was associated with elevated mortality predictions, highlighting the importance of maintaining adequate oxygen delivery in the management of TBI.

Compared to previous iterations, the length of stay in the ICU was excluded from the set of characteristics, but the model retained its predictive strength, focusing on the fact that the basic physiological and demographic variables adequately encapsulate the risk of mortality. These findings are also consistent with results from the ablation study, where key predictors like GCS score and PT yielded marked performance changes upon exclusion, validating their central role.

From a clinical perspective, SHAP values enable patient-specific transparency, allowing clinicians to visualize and interpret the most impactful drivers of model predictions. This supports precision medicine by informing customized interventions, such as increased neurologic monitoring for patients with low GCS or early hemostatic support in those with coagulopathy.

Overall, the SHAP analysis confirms that the CatBoost model integrates biologically and clinically plausible patterns, rather than functioning as a black box. Its interpretability reinforces its applicability for real-time decision support in the management of traumatic brain injury.

\section*{Discussion}
\subsection*{Summary of Existing Model Compilation}
This study developed a robust and interpretable machine learning model for predicting 30-day mortality in geriatric patients with TBI, leveraging data from the MIMIC-III critical care database. Utilizing a structured pipeline involving data imputation, hybrid feature selection, and multi-model comparison, the CatBoost algorithm achieved the highest performance, with an AUROC of 0.867 (95\% CI: 0.809–0.922), outperforming XGBoost and LightGBM. Nine features were ultimately selected, encompassing GCS score, age, oxygen saturation, prothrombin time, and vital signs—all clinically plausible predictors of short-term mortality. Notably, SHAP analysis confirmed the central role of GCS and oxygen saturation in driving predictions, reinforcing their critical importance in early TBI risk stratification. The high performance across multiple evaluation metrics and stratified cross-validation demonstrates the model’s discriminative power and potential generalizability.

The findings of this study hold significant clinical and practical value. First, the use of data within the first 24 hours of ICU admission enables early identification of patients at elevated mortality risk, providing a critical window for initiating targeted interventions, such as intensified monitoring, early palliative care discussions, or transfer to specialized neurocritical units. Second, the model’s interpretability via SHAP enhances clinician trust and supports its integration into bedside decision-making processes. For instance, when a patient presents with a low GCS score and prolonged prothrombin time, the model flags elevated mortality risk with transparent justification, potentially triggering timely resuscitative measures or anticoagulant review. Third, from a systems-level perspective, the implementation of this model can assist in triaging ICU resources more efficiently—an essential consideration given the projected rise in elderly TBI cases and the increasing pressure on critical care services. Furthermore, because the model operates on routinely collected clinical variables, it can be readily implemented across different healthcare settings without requiring additional testing or infrastructure investment.

The study has several notable strengths. It leverages a large, publicly available critical care database with granular, time-stamped data, ensuring reproducibility and wide applicability. The hybrid feature selection strategy, combining Random Forest-based importance with clinical validation, enhances both model accuracy and interpretability. Additionally, the model’s robustness was affirmed through stratified 5-fold cross-validation and statistical validation, including SHAP-based explanation and ablation studies.

\subsection*{Comparison with Prior Studies}
Several prior studies have explored prognostic modeling in geriatric TBI populations, but most rely on traditional statistical frameworks with limited predictive capacity or clinical granularity. Our findings extend and improve upon this prior work in both methodological design and outcome performance.

Fu et al. (2017) conducted a population-based analysis of TBI hospitalizations in older adults using administrative data from Canada. While their study identified advanced age, comorbidities, and injury severity as independent predictors of mortality, it did not develop or validate a formal risk prediction model. Furthermore, the lack of ICU-specific physiological data limited its applicability to acute care triage or early mortality prediction. In contrast, our model leveraged granular ICU data from MIMIC-III, including vital signs, laboratory values, and early GCS scores, all within the first 24 hours of admission. This allows for more immediate and actionable risk estimation.

Bobeff et al. (2019) developed eTBI Score to predict 30-day mortality or vegetative state in patients aged $\geq$65. Using logistic regression, their model identified key predictors such as GCS motor score (OR 0.17), comorbid organ dysfunction or malignancy (OR 2.86), platelet count $\leq 100 \times 10^9\,\text{/L}$ (OR 13.60), and \text{RDW} $\geq$ \text{14.5\%}. While the eTBI Score offered a parsimonious and clinically interpretable tool, it was developed from a relatively small sample (n=214), with limited validation and no external test cohort. Our study, in contrast, included 667 patients and utilized stratified training and testing with 5-fold cross-validation to ensure robust model generalization. 

Huang et al. (2024) evaluated the GTOS in a multicenter cohort of 5,543 older adults with moderate to severe isolated TBI. GTOS demonstrated an AUC of 0.813, with a cutoff score of 121.5 identifying patients at elevated mortality risk (OR 2.64; 95\% CI 1.93–3.61). While their model benefited from a large sample size, it did not incorporate ICU-specific physiologic features (e.g., oxygen saturation, prothrombin time), and was based on linear scoring systems. In comparison, our machine learning approach (CatBoost) captured non-linear interactions among clinical features, resulting in superior predictive accuracy and better calibration. Importantly, our inclusion of SHAP analysis provided an interpretable framework for identifying the most influential predictors—namely GCS, oxygen saturation, and prothrombin time—which aligns with known TBI pathophysiology but enhances bedside decision-making by quantifying their individual impact.

In summary, while prior TBI-focused models such as eTBI and GTOS have provided valuable foundations for risk stratification, our study advances the field by integrating a high-resolution ICU dataset, robust machine learning methodology, and interpretability tools into a unified framework that offers improved accuracy and practical clinical utility in geriatric TBI care.

\subsection*{Limitations and Future Work}
Despite these strengths, our study has limitations. It is based on retrospective, single-center data from the MIMIC-III database, which may limit external generalizability. Additionally, we restricted our input features to data available within the first 24 hours of ICU admission. While this design supports early risk stratification, incorporating time-series data could further improve predictive performance by capturing clinical deterioration trajectories. Moreover, external validation on diverse ICU datasets, including prospective and multi-center cohorts, is necessary before deployment in routine clinical practice.

In future work, we plan to evaluate temporal deep learning models (e.g., LSTM, transformers) and integrate imaging or cognitive outcomes to provide more holistic risk assessments. We also anticipate developing a user-friendly interface to integrate this model into electronic health record systems, enabling real-time clinical decision support.

\section*{Conclusion}
In this study, we developed and validated a clinically grounded, machine learning-based framework to predict 30-day mortality among geriatric patients with TBI using the MIMIC-III critical care database. A structured pipeline was employed, incorporating rigorous patient selection, Random Forest-based data imputation, and hybrid feature selection. From an initial pool of 69 candidate variables, nine clinically relevant predictors—including GCS score, prothrombin time, and oxygen saturation—were retained to ensure model interpretability and robustness.

Among the models evaluated, the CatBoost algorithm demonstrated superior discriminative performance, achieving an AUROC of 0.867 (95\% CI: 0.809–0.922), outperforming LightGBM and XGBoost. The model maintained a strong balance between sensitivity and specificity and demonstrated consistent generalizability in stratified 5-fold cross-validation. Feature ablation studies confirmed the necessity of each selected predictor, while SHAP analysis highlighted the relative importance of neurological, respiratory, and coagulation-related parameters in mortality risk stratification.

These results have meaningful clinical implications. By leveraging only early ICU data from the first 24 hours of admission, the model supports timely identification of high-risk TBI patients, enabling earlier interventions, more informed triage decisions, and improved end-of-life care planning. The model's transparency also facilitates clinical acceptance by offering interpretable predictions based on established physiological markers.

Looking forward, future work should aim to externally validate this model in multi-institutional cohorts and integrate additional modalities such as imaging data or frailty indices. Ultimately, this framework lays the foundation for real-time, bedside risk prediction tools that can be seamlessly incorporated into ICU workflows, helping clinicians optimize care delivery and allocate resources effectively in the growing elderly TBI population.

\section*{Acknowledgments}
 Y.S. conceptualized the study, developed the methodological framework, conducted the experiments, performed data analysis, and drafted the original manuscript. S.C., J.F., and L.S. contributed to data preprocessing, model development, and manuscript preparation. E.P., K.A., and G.P. provided critical feedback on the study design and interpretation of results. M.P. supervised the project, coordinated research efforts, and provided strategic guidance throughout the study. All authors reviewed and approved the final version of the manuscript.

The authors would also like to acknowledge the Laboratory for Computational Physiology at the Massachusetts Institute of Technology for maintaining the MIMIC-III database.


%
%
%

\bibliography{referrences}

\end{document}